\documentclass[twocolumn,aps,prl]{revtex4-1}

\usepackage{graphicx}
\begin{document}

\title{Robust triboelectric charging of identical balloons of different radii}

\author{Francisco Vera}
\email{francisco.vera.ucv@gmail.com}
\author{Rodrigo Rivera}
\author{Manuel Ortiz}
\affiliation{Instituto de F\'{\i}sica, Pontificia Universidad Cat\'{o}lica de Valpara\'{\i}so.\\
Av. Universidad 330 Curauma, Valpara\'{\i}so, Chile.}
\author{Francisco Antonio Horta-Rangel}
\affiliation{Divisi\'{o}n de Ingenier\'{\i}as del C.G., Universidad de Guanajuato, M\'{e}xico}


\date{\today}
\maketitle

Electrification by rubbing different materials is a well known phenomenon with a history that begun more than five centuries BC. However, simple experiments can lead to contradictory or inconsistent results and the history of this phenomena is plagued with non-intuitive results. For example, triboelectric charging by rubbing identical materials is possible. In this work we want to highlight some historical aspects of triboelectricity that could enrich the discussion of electrostatics in an undergraduate physics course. We will focus on the effect of strain on the triboelectric properties of a sample, which we think is not well known to physics teachers. We will show that it is possible to obtain robust polarities by rubbing identical rubber balloons of different radii and we will also show that this charging method can be very useful in introductory physics courses.

In practical terms, one important idea was stablished by J. C. Whilke in 1757, who proposed to arrange materials in a triboelectric series. When two material in the series are rubbed, the relative order of the materials in the series determine which one becomes positively charged and which one becomes negatively charged. Thus, a given material can acquire any sign of charge depending on which material is rubbed against it.  The phenomenon of triboelectric charging is so complex that a basic understanding of the physics behind it is not clear nowadays, even in matters so basic as to whether electrons, ions or bulk material are the particles transferred that are responsible of the charging process. Besides, multiple factors have incidence on the result of the triboelectric charging process, among them: the chemical nature of the surface, its morphology, the interaction with the environment and even the past history of the samples. A simple review of the history of static electricity and the triboelectric series can be read in the review written in 2019 by Lacks and Shinbrot\cite{Lacks-2019} and in the introduction written by Pionteck for the Handbook of Antistatics\cite{Pionteck-2016}.

Recently Zou et al.\cite{Zou-2019}, working under well controlled experimental conditions, built a beautiful quantitative triboelectric series. Although the results are attractively simple, they can lead to the unrealistic conclusion that it is possible to  obtain a robust tribolectric table, when it is known that a simple physical change in a sample can considerably alter its triboelectric properties.

A very simple experiment that initiated the discussion of the effects of strain in triboelectricity was described in 1910 by Jamieson\cite{Jamieson-1910}. When two narrow strips of celluloid are pulled between the fingers, each strip surface is electrified with a different polarity. This experiment is similar to the two strip experiment used in many inquiry based activities to teach electrostatics\cite{McDermott,Mazur}, but we are not aware of a historical connection between them. In the experiment reported in 1910, the charging of the strips was a robust fact, but the polarities of the strips seemed to be arbitrary. In his note, Jamieson reports that it was Mr. M. McCallum Fairgrieve who noted that the sign of each strip was related to the direction of the bending of the contacting surfaces of the strips. Thus, the important factor was whether each surface was under compression or under tension, the compressed celluloid surface being the one that turned always negatively charged. 
The triboelectric charging of identical materials was studied by Shaw in 1926\cite{Shaw-1926}. He rubbed two identical ebonite rods, keeping one of them fixed (the “rubbed” rod) and moving the other one (the “rubber” rod with a small contact area) down the first rod. Shaw observed that consistently the rubber rod acquired negative charge and the rubbed rod acquired positive charge, with a similar result in other materials. Surprisingly, if he continued rubbing the rods beyond this point, the rubber rod gradually decreased its negative charge, became neutral and then charged increasingly positive. The rods could be returned to their initial state by boiling them in water. Shaw explained these effects by the strain produced in the rods by the asymmetric way of rubbing them. Since then, several experiments have established that strained materials change its triboelectric behavior.
One of the experiments used by Shinbrot et. al.\cite{Shinbrot-2008} to study tribocharging of similar materials is very simple: it consists in rubbing two identical balloons (of the same radius). It is found that a symmetry breaking occurs since the balloons become charged with charges of different sign and same magnitude, but the choice of balloon polarity seems to be random. Another experiment more easily controlled uses two identical polycarbonate disks rolled against each other and similar results are obtained.

In 2012 Sow et. al.\cite{Sow-2012} used another setup to study the effect of material strain on contact electrification of different materials. This time a latex rubber balloon was put in contact with a polytetrafluoroethylene (PTFE or teflon) surface whose charge was measured. The sign of the charge in the PTFE depended on whether the balloon was deflated or was given the shape of an inflated (and therefore strained) balloon. Since this setup could not be performed in a controlled manner, the same research group\cite{Sow-Lacks-2012} performed another version of the experiment where a PTFE sphere was repeatedly contacted with a strained flat latex surface. When the strain in the latex surface was larger than a certain value, the sign of the surface potential changed sign towards positive values. The authors of these works concluded that a rigorous ordering of materials in a triboelectric series appears to be unrealistic.

In the past few years we have been building Van de Graaff generators using unconventional materials for the rollers. One of our designs included identical rollers that charged the dome with random polarities. While discussing the physics behind this setup, one of our teaching assistants rubbed two balloons of the same material and was surprised of the resulting different polarities in them. Further experimentation lead us to discover that rubbing two identical balloons inflated to different sizes produced a robust polarity sign that was correlated with the relative size of the balloons. The main result is that the larger (more stressed) balloon gets negatively charged.

\begin{figure}
 \centering
\includegraphics[width=1.0\columnwidth]{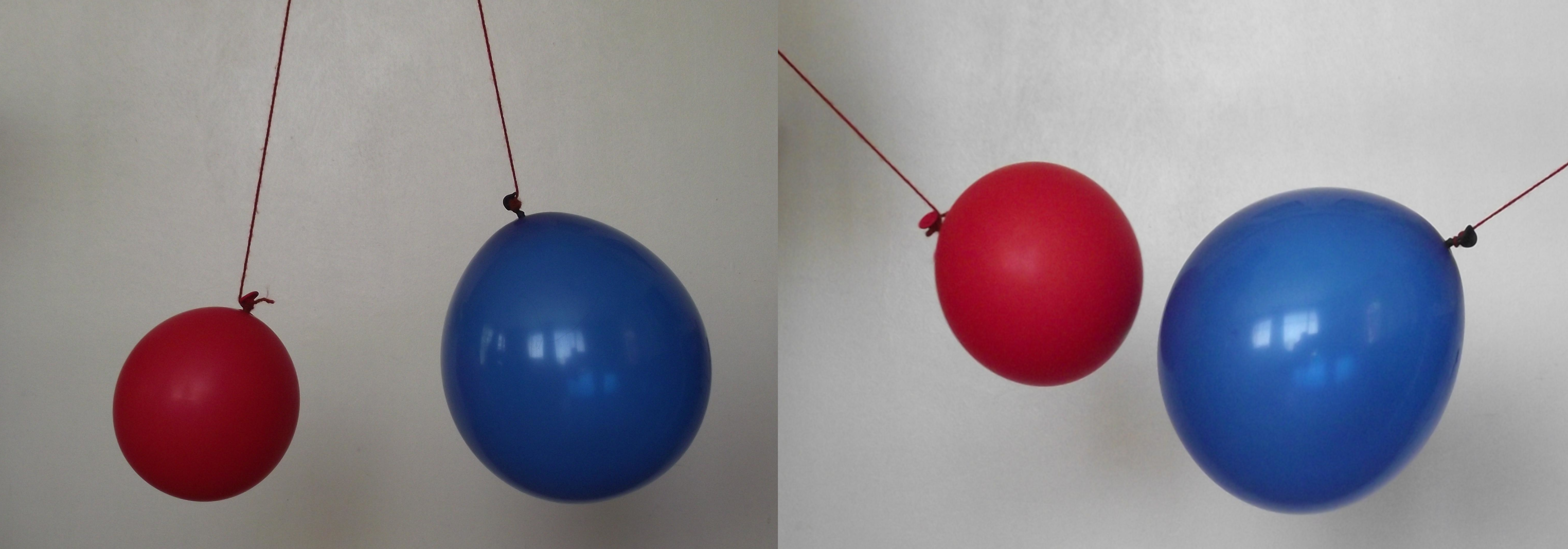}
 \caption{ Two balloons of different sizes. Left: both balloons charged negative when initially uncharged and rubbed against hair. Right: balloons charged with different polarities when rubbed against each other.}
\end{figure}
We have made experiments using several brands of rubber balloons, and we also used small water balloons. In every case we inflated identical balloons to different radii and the predicted polarities where obtained. In a typical experiment at the laboratory we used balloons of diameters 10 cm, 20 cm and 25 cm, the polarity of the balloons was measured using some of the polarity detectors for physics demonstrations that we reported previously\cite{Capacitors,Tiburon}. The magnitude of the charge in the balloons depended to some extent on the kind of balloons used, but in all cases the polarity of the charges was consistent as described above.

In another setup, we begun with three balloons initially uncharged, then they were rubbed against hair and after all balloons become negatively charged we rubbed any pair of balloons. For any pair of balloons in this experiment, the larger balloon gets negatively charged and the initial polarities change accordingly.

The robust charging of two initially discharged balloons of different radii offers for example a simple alternative to the simple experiment proposed by Rueckner in 2007\cite{Rueckner} to demonstrate charge conservation. To the best of our knowledge, the robust charging of two rubbed balloons has not been noted before and it could be used in schools and undergraduate labs to provide a simple mechanism to obtain specific signs of charges for a variety of experiments.

\begin{acknowledgments}
We would like to acknowledge financial support from Fondo Nacional de Desarrollo Cient\'{\i}fico y Tecnol\'{o}gico, FONDECYT Project 1181782.
\end{acknowledgments}

\end{document}